\documentclass{article}
\usepackage{ijcai20}

\usepackage{times}

\usepackage{soul}
\usepackage{url}
\usepackage[hidelinks]{hyperref}
\usepackage[utf8]{inputenc}
\usepackage[small]{caption}
\usepackage{graphicx}
\usepackage{amsmath}
\usepackage{booktabs}
\title{IIT Kanpur Consulting Group: Using Machine Learning and Management Consulting for Social Good}

\author{
Tushar Goswamy$^1$\footnote{Contact Author}\footnote{All the authors contributed equally}\and
Vatsalya Tandon$^1$\footnotemark[2]\and
Naishadh Parmar$^{1}$\footnotemark[2]\and
Raunak Shah$^1$\footnotemark[2]\and
Ayush Gupta$^1$\footnotemark[2] \\
\affiliations
$^1$Indian Institute of Technology (IIT) Kanpur, Kanpur, India\\
\emails
\{tgoswamy, vatsalt, naishadh, raunaks, ayushgup\}@iitk.ac.in
}

\begin{document}

\maketitle

\begin{abstract}
The IIT Kanpur Consulting Group is one of the pioneering research groups in India which focuses on the applications of Machine Learning and Strategy Consulting for social good. The group has been working since 2018 to help social organizations, nonprofits, and government entities in India leverage better insights from their data, with a special emphasis on the healthcare, environmental, and agriculture sectors. The group has worked on critical social problems which India is facing including Polio recurrence, COVID-19, air pollution and agricultural crop damage. This position paper summarises the focus areas and relevant projects which the group has worked on since its establishment, and also highlights the group's plans for using machine learning to address social problems during the COVID-19 crisis. 
\end{abstract}

\section{Introduction}

The IITK Consulting Group was established in 2018 at IIT Kanpur, to work on problems which need to be analyzed in a management consulting setting to arrive at the key issues and then address them using artificial intelligence based solutions. The group also places a strong emphasis on research-based projects. The group's focus has been to work with social organizations, nonprofits, and government entities to help them leverage better insights from their data to create a strong social impact. 

When the group was set up, two kinds of organizations existed - firms like Google and Microsoft who worked on pure machine learning projects with a focus on research and algorithms, and management consulting firms like McKinsey \& Company, the Boston Consulting Group and Bain \& Company who work on strategic solutions for their clients' problems. We identified the combined potential of analyzing and narrowing down a problem statement to salient pain points using management consulting, and using machine learning to draw significant insights from client's data to address those pain points precisely, thus making the most effective use of time, workforce, resources and, most importantly, data to drive maximum impact. Some technology consulting firms such as Mu Sigma are already implementing this model but for businesses and not social organizations. We wanted to test this model on socially relevant problems, especially in India. Thus, we work with organizations who face specific problems in their operations and have the relevant data but do not know how to leverage that data to solve their problem, and cannot afford the services of firms like Mu Sigma. 

A typical project involves helping the partner organization identify the scope for machine learning in solving the problem they are facing, as well as exploring areas where machine learning can help improve the efficiency of their operations. This step is followed by building the actual pipeline and submitting it to the organization to test the model and share feedback for improvement. Since we focus more on research-oriented projects with these organizations, the projects usually culminate with a research paper outlining our approach and results so that researchers across the globe can use our work and improve upon it. We have currently undertaken projects in the following sectors, and are gradually expanding our domain into new sectors:

\begin{enumerate}
    \item Healthcare
    \item Environment
    \item Agriculture
\end{enumerate}

The group primarily functions under the guidance of the Department of Computer Science \& Engineering, IIT Kanpur, and we collaborate with researchers from other departments when domain expertise is required for a particular project. 

\section{Related Work}

Several companies have started their own AI for Social Good division, and some research labs affiliated with Universities are also focussing on research in this area. On the corporate side, Google has a separate division called AI for Social Good under Google AI research, which has done projects such as using AI for flood forecasting in Bihar, India \cite{47651}, and predicting the risk of cardiovascular diseases of a patient using his/her retinal images \cite{46425}. Microsoft Research India has set up a research group called Center for Societal Impact through Cloud \& Artificial Intelligence, which has done projects like creating a mental health app to help distressed people with mental health concerns \cite{johnson2020barriers}. 

As far as research labs at Universities are concerned, some notable groups include the Center for Research on Computation \& Society, Harvard University, Data Science for Social Good Fellowship which was started at the University of Chicago but now operates out of the Carnegie Mellon University, and the AI for Humanity group at the Montreal Institute for Learning Algorithms (MILA). In India, the Wadhwani Institute for Artificial Intelligence is a nonprofit research institute that works on developing AI solutions for social good.    

\section{Operations}
We first identify a social organization that might be facing a problem, followed by networking with representatives from the organization. After networking with them, we work on understanding the problems they are facing and where we can fit in to ameliorate their situation. All these projects are undertaken on a  pro-bono basis. After getting a precise specification of the problem to be solved, we recruit students from IIT Kanpur with prior experience in either machine learning or in the domain that the particular problem is from, and the project is then assigned a leader who is usually the person who contacted the client organization in the first place. The leader formulates a plan of action and distributes work among the recruited team members and stays in constant communication with the client organization regarding the project's developments. After reaching significant milestones, the entire project team meets with the client with the results. A discussion on the results is held, after which the plan of further action is revised. When the project ends, the final solution is submitted to the client organization.

\section{Focus Areas \& Project Overview}

\subsection{Healthcare}

\subsubsection{1. Risk Mitigation Framework to combat the Recurrence of Polio Disease in India}

The group's pilot project was in partnership with the India office of Rotary International, to classify regions in India based on their risk of Polio recurrence. It has been six years since India witnessed its last reported case of Polio, but its neighbors Pakistan and Afghanistan are among the three countries in the world where the disease has not been eradicated yet. These countries pose an imminent risk of transmission of the poliovirus into India. We used a Feature-weighted Fuzzy C-Means Clustering framework to classify cities into three risk-regions: High, Moderate, and Low Risk. 

\textbf{Impact:} The clustering model's results were used to devise a resource allocation plan as well as a cost estimate for Rotary India to help with its mitigation efforts towards reducing the risk of Polio disease in India. Through this research, we demonstrated the application of unsupervised learning for risk assessment of regions towards diseases that have been eradicated but pose a threat of reemergence. To the best of our knowledge, this is the first and the only research that does a quantitative risk assessment of a region towards poliovirus' recurrence using an unsupervised learning model. This work has been presented at the World Health Organisation's Regional Office for Southeast Asia and at the Student Research Seminar at The University of Tokyo. 

\subsubsection{2. AI-based Hospital Preparedness Monitoring system towards COVID-19 for Southeast Asia}

The COVID-19 pandemic has emerged as the defining global crisis of our time, and countries have endured the brunt of it. Southeast Asian countries have an extremely dense population and a poor health infrastructure, which makes them prone to the collapse of the public health care systems \cite{Coker2011}. The project is being pursued in collaboration with the Health Emergencies Program Team, WHO Regional Office for Southeast Asia. We established contact with the WHO team and took cognizance of their concerns about the non-availability of reliable hospitalization data, which did not reflect the real stress at the health care facilities which the general public was facing. We are currently working on a COVID-19 monitoring and response system, to identify the surge in the volume of patients at hospitals and healthcare centers in select countries (India, Bangladesh and Indonesia). 

We have developed a proxy to indicate the burden on health facilities by using Twitter data for signals. The system will help the authorities in developing appropriate resource planning measures. Our research aims to identify overcrowding incidents at hospitals, shortage of critical equipment like ventilators, and lack of available ICU units. This can help understand the medical preparedness levels of these countries' health facilities and the burden on their hospitals as the pandemic spreads. The project involves a machine learning pipeline consisting of scraping historical tweets at a granular level, processing them using Natural Language Processing techniques, and finally calculating the signals and applying appropriate smoothing functions. Neural translation models have also been used to account for the usage of regional language. This is followed by developing a robust alert system, where multiple time series based methods were experimented upon to build a model corroborated with our empirical evidence. 

\textbf{Impact:} The model was developed and deployed for preliminary usage by the WHO team, and has yielded accurate results for states in India. We are working on validating the model for the remaining countries so that it can serve as a reliable surveillance tool for researchers to monitor the burden on hospitals, and for authorities to take timely resource planning measures.

\subsection{Environment}

\subsubsection{1. Deep Mixture Model for Air Quality Forecasting in Delhi-NCR (National Capital Region)}

In recent years, there has been a growing concern regarding air quality among researchers and citizens alike. There is an urgent need to improve our understanding of air quality trends in megacities. Delhi National Capital Region (Delhi-NCR) specifically had the 3rd highest (Particulate Matter) PM10 levels among 39 megacities worldwide in the years 2011-2015. Pollution control boards and government organizations can implement control measures with greater foresight if an efficient air quality forecasting system is already in place.

To address this problem, we partnered with the Ministry of Environment, Government of India's National Clean Air Program, to develop a deep learning-based mixture model for forecasting PM2.5 levels. The model predicts the PM2.5 levels and its probability distribution at each of the 13 sensor locations in the city up to 48 hours into the future using AI techniques. The inputs to the models use climatic as well as pollutant-based features such as temperature, humidity, visibility, wind speed, wind direction, PM1, PM10, and PM2.5 concentrations taken up to 6 hours in the past.

\textbf{Results:} Our work has made an impact in the scientific community. According to our literature review, a probabilistic prediction of up to 48 hours has never been made before in this city. A mixture-based methodology is also novel applied to the pollution forecast of this city. Our work was presented by the supervising professors in the Workshop on Low-Cost Sensor-based Real-time Air Quality Monitoring with advances in IoT, Machine Learning and AI, New Delhi. Our work was accepted for a poster presentation in the International Conference On Challenges Of Air Quality In Global Megacities (CAGMe-2020). Manuscript of the paper to be published in other reputed journals is under development. 

\subsubsection{2. Improving the GFS weather forecasts for improved Risk Management of Farmers in Agriculture}

The project has been undertaken to help Weather Risk Management Services Private Limited, a pioneering smart farming startup. They leverage data, technology, and financial services to provide holistic solutions that enhance agricultural productivity and secure farmers' income in an environmentally sustainable manner. The startup wanted to provide solutions to their farmers with greater confidence, so we provided a management consulting solution to them after studying the framework they followed for providing those solutions. Our key findings were that their recommendations to farmers are dependent on the accuracy of their weather forecasts for the agricultural fields and that they were using the forecasts provided by the Global Forecasting System (GFS) \cite{cisl_rda_ds084}. We identified that improving upon the GFS forecasts of temperature and rainfall for the agricultural fields will give them a significant edge in the solutions they provide. 
We developed time series forecasting models that improved upon the temperature forecasts and brought them under 2\% MAPE as per the organization's requirements. We are currently working on a spatio-temporal model to predict unseasonal rainfall in non-monsoon months and rainfall anomalies in monsoon months for their agricultural fields. 

\textbf{Impact:} The project directly impacts the core functioning of the organization in terms of the recommendations and the solutions they provide to their farmers. Indirectly, our solution assists farmers to minimise any losses which they might incur due to anomalous weather conditions and to save their crops from getting destroyed. This ensures that food supplies are not hampered for the general population. 

\section{Future Projects}

\subsection{AI-based Land Cover change and Deforestation Prediction}

We have partnered with Moja Global, an organization that provides tools for estimating emissions and removals of greenhouse gases (GHG) from the land sector. Agriculture, forestry, and other types of land use account for 23\% of global GHG emissions. To reduce these emissions, they need to be quantified accurately. Deforestation is one of the primary sources of GHG emissions from the land. Reductions of GHG emissions from land are measured against a business-as-usual projection.

This project proposes developing a machine learning-based model for forecasting deforestation to improve upon the organization's existing method. We will use models based on Convolutional Neural Networks to predict how likely it is for a particular pixel in the landscape to be deforested over the projection period. The predicted deforestation for each pixel can be aggregated to obtain expected deforestation for a project area, a sub-national administrative region, or a whole country.

\subsection{AI for Clean India}

India generates 62 million tonnes of waste annually, with an average annual growth rate of 4\%. Out of this, less than 60\% is collected and only 15\% finally gets processed. One of the major challenges that the Municipal Department faces is identifying areas where a garbage dump has developed and can lead to sanitation issues for the people living nearby. Another issue in the domain of waste management is segregating the garbage and identifying elements which can be recycled.

The aim of this project is to utilise technology to posit a low-cost and easy-to-implement solution. We would help in optimising the waste collection drives by creating a map of unchecked garbage dumps. This would work on crowd-sourced data in addition to aerial images of the city captured using surveillance drones. The application would be scalable, and after initial testing in the local neighbourhoods, it would be extended across the city.

\section{Impact and Future Plan}

We have collaborated with organizations and startups that work at the frontlines of solving social problems in the domain of healthcare, environment and agriculture, in order to aid them in a data-driven manner and drive maximum impact in the social problems they solve. Our solutions have helped not only these social organizations and the communities that they impact, but also the scientific world in terms of the research that we perform as we solve the problems of these organizations.

In the future, we plan to partner with organizations in allied domains such as water, education, economy, and public policy. There is enormous scope for extracting useful and actionable insights from data in these sectors, and our group envisions to empower organisations in India and abroad with the right expertise to create a better, brighter, and more hopeful future for all of us.

\bibliographystyle{named}
\bibliography{main}

\end{document}